\PassOptionsToPackage{unicode}{hyperref}
\PassOptionsToPackage{hyphens}{url}
\PassOptionsToPackage{dvipsnames,svgnames,x11names}{xcolor}
\documentclass[
  12pt]{article}

\usepackage{amsmath,amssymb}
\usepackage{iftex}
\ifPDFTeX
  \usepackage[T1]{fontenc}
  \usepackage[page]{appendix}
  \usepackage[utf8]{inputenc}
  \usepackage{textcomp} 
\else 
  \usepackage{unicode-math}
  \defaultfontfeatures{Scale=MatchLowercase}
  \defaultfontfeatures[\rmfamily]{Ligatures=TeX,Scale=1}
\fi
\usepackage{lmodern}
\ifPDFTeX\else  
\fi
\IfFileExists{upquote.sty}{\usepackage{upquote}}{}
\IfFileExists{microtype.sty}{
  \usepackage[]{microtype}
  \UseMicrotypeSet[protrusion]{basicmath} 
}{}
\makeatletter
\@ifundefined{KOMAClassName}{
  \IfFileExists{parskip.sty}{%
    \usepackage{parskip}
  }{
    \setlength{\parindent}{0pt}
    \setlength{\parskip}{6pt plus 2pt minus 1pt}}
}{
  \KOMAoptions{parskip=half}}
\makeatother
\usepackage{xcolor}
\makeatletter
\ifx\paragraph\undefined\else
  \let\oldparagraph\paragraph
  \renewcommand{\paragraph}{
    \@ifstar
      \xxxParagraphStar
      \xxxParagraphNoStar
  }
  \newcommand{\xxxParagraphStar}[1]{\oldparagraph*{#1}\mbox{}}
  \newcommand{\xxxParagraphNoStar}[1]{\oldparagraph{#1}\mbox{}}
\fi
\ifx\subparagraph\undefined\else
  \let\oldsubparagraph\subparagraph
  \renewcommand{\subparagraph}{
    \@ifstar
      \xxxSubParagraphStar
      \xxxSubParagraphNoStar
  }
  \newcommand{\xxxSubParagraphStar}[1]{\oldsubparagraph*{#1}\mbox{}}
  \newcommand{\xxxSubParagraphNoStar}[1]{\oldsubparagraph{#1}\mbox{}}
\fi
\makeatother

\usepackage{longtable,booktabs,array}
\usepackage{calc} 
\usepackage{etoolbox}
\makeatletter
\patchcmd\longtable{\par}{\if@noskipsec\mbox{}\fi\par}{}{}
\makeatother
\IfFileExists{footnotehyper.sty}{\usepackage{footnotehyper}}{\usepackage{footnote}}
\makesavenoteenv{longtable}
\usepackage{graphicx}
\makeatletter
\def\maxwidth{\ifdim\Gin@nat@width>\linewidth\linewidth\else\Gin@nat@width\fi}
\def\maxheight{\ifdim\Gin@nat@height>\textheight\textheight\else\Gin@nat@height\fi}
\makeatother
\setkeys{Gin}{width=\maxwidth,height=\maxheight,keepaspectratio}
\makeatletter
\def\fps@figure{htbp}
\makeatother

\makeatletter
\@ifpackageloaded{caption}{}{\usepackage{caption}}
\AtBeginDocument{%
\ifdefined\contentsname
  \renewcommand*\contentsname{Table of contents}
\else
  \newcommand\contentsname{Table of contents}
\fi
\ifdefined\listfigurename
  \renewcommand*\listfigurename{List of Figures}
\else
  \newcommand\listfigurename{List of Figures}
\fi
\ifdefined\listtablename
  \renewcommand*\listtablename{List of Tables}
\else
  \newcommand\listtablename{List of Tables}
\fi
\ifdefined\figurename
  \renewcommand*\figurename{Figure}
\else
  \newcommand\figurename{Figure}
\fi
\ifdefined\tablename
  \renewcommand*\tablename{Table}
\else
  \newcommand\tablename{Table}
\fi
}
\@ifpackageloaded{float}{}{\usepackage{float}}
\floatstyle{ruled}
\@ifundefined{c@chapter}{\newfloat{codelisting}{h}{lop}}{\newfloat{codelisting}{h}{lop}[chapter]}
\floatname{codelisting}{Listing}

\makeatother
\makeatletter
\@ifpackageloaded{caption}{}{\usepackage{caption}}
\@ifpackageloaded{subcaption}{}{\usepackage{subcaption}}
\makeatother

\ifLuaTeX
  \usepackage{selnolig}  
\fi
\usepackage[]{natbib}
\usepackage{bookmark}

\IfFileExists{xurl.sty}{\usepackage{xurl}}{} 
\hypersetup{
  pdftitle={Title},
  pdfauthor={Author 1; Author 2},
  pdfkeywords={3 to 6 keywords, that do not appear in the title},
  colorlinks=true,
  linkcolor={blue},
  filecolor={Maroon},
  citecolor={Blue},
  urlcolor={Blue},
  pdfcreator={LaTeX via pandoc}}

\newcommand{\anon}{1}

\begin{document}

\def\spacingset#1{\renewcommand{\baselinestretch}%
{#1}\small\normalsize} \spacingset{1}


\if1\anon
{
  \title{\bf Smoothly Time-Varying Continuous Time Markov Chains in Phylogenetics}
  \author{Pratyusa Datta\\
    Department of Biostatistics, University of California Los Angeles\\
    Philippe Lemey \\
    Department of Microbiology, Immunology and Transplantation, \\Rega Institute, KU Leuven\\
    Marc A. Suchard\thanks{
        Corresponding author: msuchard@ucla.edu}\hspace{.2cm}\\
    Department of Biostatistics, University of California Los Angeles
    }
   \date{}
  \maketitle
} \fi

\if0\anon
{
  \bigskip
  \bigskip
  \bigskip
  \begin{center}
    {\LARGE\bf Smoothly Time-Varying Continuous Time Markov Chains in Phylogenetics}
      \author{Pratyusa Datta}
\end{center}
  \medskip
} \fi

\bigskip
\begin{abstract}
The dependence of evolutionary rate estimates on the timeframe of sampling poses a fundamental challenge for reconstructing evolutionary histories from molecular sequence data, which is central to evolutionary biology and infectious disease research.
We present a novel and flexible approach to accommodate time-varying evolutionary rates by modeling the sequence substitution process using inhomogeneous continuous-time Markov chains (ICTMCs) acting along the branches of the phylogeny, and parameterizing the log transformed rate as a smooth function of time using a cubic B-spline basis expansion. 
Following the parlance of phylogenetics that refers to rates of molecular substitutions as molecular clocks, we call this a spline clock model.
Integrals of the rate function over all branches, required for likelihood evaluation, are approximated efficiently using Gauss-Legendre quadrature, and smoothness is enforced by assigning a Gaussian Markov random field prior to the spline coefficients.
Through a simulation study, we demonstrate that the spline clock model recovers the true time-varying rates more accurately and with tighter credible intervals than competing clock models. 
We apply the spline clock model to examine the evolutionary rate of foamy virus and the rate of spatial diffusion of SARS-CoV-2 across Europe, recovering strong time-varying signal in both settings.
\end{abstract}


\newpage
\spacingset{1.8} 

\section{Introduction}
\subsection{Background}
Reconstructing time-scaled evolutionary histories from molecular sequence data plays a key role in evolutionary biology and infectious disease research.
Phylogenetic trees represent evolutionary histories and describe how different biological entities are related through their branching structure, with the branches themselves reflecting the accumulation of genetic substitutions over time.
Calibrating the phylogenetic tree by converting branch lengths into calendar time often relies on external information about the geological age of one or multiple internal nodes from fossil records  \citep{YangRannala2005} or sequences collected at different time points \citep{Rambaut2000}.
Standard likelihood-based approaches to phylogenetic inference model the observed molecular sequences as having evolved under a homogeneous continuous-time Markov chain (CTMC) acting along the branches of the phylogeny. 
In the context of divergence time estimation, the CTMC is rescaled through a molecular clock model that assigns an evolutionary rate to each branch, where the evolutionary rate is defined as the expected number of substitutions per site per unit time.
The simplest molecular clock model is the strict clock \citep{Zuckerland1965} that constrains all branches to share a single common rate. 
Relaxed clock models \citep{Yoder2000, Kishino2001, Yang2003, Drummond2006, Drummond2010} depart from this assumption by allowing the evolutionary rate to vary across branches or lineages, differing in the specific distributional or correlation assumptions imposed on this variation.

\subsection{Challenges and Existing Work}
Phylogenetic methods are widely used in infectious disease research to reconstruct the transmission history of pathogens and identify sources of outbreaks \citep{Pybus2009}.
Many important pathogens, particularly RNA viruses, evolve rapidly with their genomes accumulating appreciable genetic change over periods ranging from decades to even weeks or days. 
This enables researchers to collect viral genome sequences in real time throughout the course of an outbreak \citep{dudas2016}, and to use the sampling dates of serially collected sequences as direct calibrations for the molecular clock \citep{Rambaut2000, Drummond2002}. 
However, multiple studies have demonstrated that for several such rapidly evolving viruses, evolutionary rate estimates depend on the time scale of sampling \citep{Duchene2014, Aiewsakun2016}.
Using foamy virus as a case study, \cite{aiewsakun2015b} demonstrated that the time dependent rate phenomenon is likely responsible for the discrepancy between short-term and long-term rate estimates, and that a simple power-law rate decay model best describes the underlying rate dynamics.
\cite{Membrebe2019} developed an epoch modeling approach, borrowing from \citet{Bielejec2014} and incorporating a power-law relationship between rate and time, and demonstrated strong dependence between the evolutionary rate and time for foamy virus and lentivirus.
Building on these insights, \cite{Datta2025} modeled the sequence substitution process along the unobserved phylogeny using inhomogeneous continuous time Markov chains (ICTMCs) to accommodate the change in the evolutionary rate through time.
Under the assumption that the time-varying rate function scales all elements of the infinitesimal generator matrix equally and is positive and integrable with respect to time \citep{Rindos, Kailath1980, Fortmann1977}, the transition probability matrix of the ICTMC associated with each branch reduces to a function of the integral of the rate over that branch.
Evaluating the observed data likelihood requires computing the transition probabilities of the ICTMCs acting along all $2N - 2$ branches of the phylogeny, where $N$ denotes the number of tip nodes and each tip node represents a sampled molecular sequence. 
\cite{Datta2025} developed the polyepoch clock model where the rate is modelled as a positive, piecewise constant function of time, making the computation of the transition probabilities analytically tractable and inexpensive and thereby  bypassing the computational challenges that would rise under a fully nonparametric framework.
\cite{Datta2025} demonstrated substantial variation in the evolutionary rate of multiple rapidly evolving viruses under the polyepoch clock model.
While this model allows the rate to vary flexibly across epochs, the  discretization of time imposes discontinuities.  
Consequently, the resulting estimates can exhibit spurious local oscillations mixed with genuine temporal signal, particularly in regions of dense taxon sampling.

\subsection{Our Contribution}
In this paper, we present the spline clock model, where we model the sequence substitution process using ICTMCs and propose a novel approach by modeling the log transformed  rate as a smooth function of time using a B-spline basis expansion.
This yields estimates that vary continuously and smoothly over the entire temporal domain.
Spline-based methods occupy a central place in semiparametric and nonparametric function estimation, owing to their ability to adaptively capture complex functional forms while enforcing smoothness \citep{Eubank1999, Wahba1990, GreenSilverman1994}.
In this work, we use cubic B-splines, which have attractive smoothness properties.
The transition probabilities of the ICTMCs under the spline clock model involve integrals of the exponentiated B-spline function that needs to be approximated numerically.
Owing to the smooth nature of the spline parameterization, the transition probabilities can be approximated efficiently and with high accuracy using Gauss-Legendre quadrature \citep{Gauss1814, Press2007}.
We place a Gaussian Markov random field (GMRF) prior \citep{rueheld} on the spline coefficients that  prevents overfitting by penalizing differences between adjacent coefficients  \citep{LangBrezger2004} and  naturally accommodates uncertainty in the degree of smoothing through a precision parameter that is assigned a hyperprior and estimated from the data. 
We demonstrate that the spline clock model yields smooth and accurate estimates of time-varying evolutionary rates through a simulation study.
We then apply the model to examine the temporal variation in the evolutionary rate of the foamy virus and the rate of spatial diffusion of SARS-CoV-2 across Europe in 2020.
Our findings reveal strong time varying patterns in the rate estimates obtained for both exemplars.

\section{Methods}\label{sec:Methods}

\newcommand\data{\mathbf{Y}}
\newcommand{\datum}{Y}
\newcommand\columns{C}
\newcommand\column{c}
\newcommand\states{S}
\newcommand\state{s}
\newcommand\numtips{N}
\newcommand\branchLengths{T}
\newcommand\branchLength{b}
\newcommand\terminalBranches{\varepsilon}
\newcommand\internalBranches{\mathcal{I}}
\newcommand\phylogeny{{\cal F}}
\newcommand\branch{b}
\newcommand\rates{R}
\newcommand\rate{r}
\newcommand{\siteSpecificRate}{\gamma}
\newcommand\postorderPartial{\mathbf{p}}
\newcommand\postorderElement{p}
\newcommand\preorderPartial{\mathbf{q}}
\newcommand\preorderElement{q}
\newcommand{\probabilityMatrix}[2]{\mathbf{P}^{(#1)} \hspace{-0.2em} \left( #2 \right)}
\newcommand{\probabilityEntry}[4]{P^{(#1)}_{#3 #4} \hspace{-0.2em} \left( #2 \right)}
\newcommand\rateMatrix{\mathbf{Q}}
\newcommand\disjointSet{Y}
\newcommand{\below}[1]{\mathbf{Y}_{\lfloor #1 \rfloor}}
\newcommand{\abbove}[1]{\mathbf{Y}_{\lceil #1 \rceil}}
\newcommand{\transpose}{'}
\newcommand{\estate}{t}
\newcommand{\parent}{k}
\newcommand{\sibling}{j}
\newcommand{\probability}[1]{\mathbb P \hspace{-0.2em} \left( #1 \right)}
\newcommand{\rootDistribution}{\boldsymbol{\pi}}
\newcommand{\given}{\,|\,}
\newcommand{\pa}[1]{{\mathrm{pa}(#1)}}

\subsection{Problem Setup}

Given $N$ aligned molecular sequences, we consider an unobserved phylogenetic tree $\mathcal{F}$ with $N$ tip nodes and $N - 1$ internal nodes, that describes the evolutionary history shared by the $N$ observed sequences.
Each branch of $\mathcal{F}$ joins a parent node to its child node, where the parent node is closer to the root. 
The tip nodes are indexed $1, \dots, N$ and the internal nodes are indexed $N + 1, \dots, 2N - 1$, where the root node is indexed $2N - 1$.
Sequences are observed exclusively at the tip nodes and are unobserved at all internal nodes.
The parent of node $i$ is denoted with $\pa{i}$.
The real time associated with node $i$ is denoted with $t_i$ and the length of the branch connecting node $i$ to $\pa{i}$ is denoted with $b_i$, where branch length is measured in expected number of substitutions along the branch per site, and we label the branch length by the index of the child node.
Following \cite{Datta2025}, we model the sequence alignment as arising from conditionally independent ICTMCs acting along the branches of $\mathcal{F}$ to accommodate the change in evolutionary rate through time.
The state-space of the ICTMC is denoted with $\mathcal{S}$ and $S$ is the number of possible states in $\mathcal{S}$ (e.g. $S = 4$ for nucleotide substitution models).
The number of aligned sites is denoted with $C$.
We denote observed (at tips) and latent (at internal nodes) discrete evolutionary characters at site $c$ of the $i$-th node with $Y_{i,c}$, where $Y_{i,c} \in S$, for $i = 1, \dots, 2N - 1$, and $c = 1, \dots, C$. 
The observed data $\mathbf{Y} = (\mathbf{Y}_1, \dots, \mathbf{Y}_{N})$ consists of columns  $\mathbf{Y}_i = (Y_{i,1}, \dots, Y_{i,C})\transpose$ that are conditionally independent, for $i  = 1, \dots, N$. 
Across-site rate variation is commonly accommodated via a discretized-Gamma prior on site-specific rates, or through more complex substitution processes \citep{Gill:2025aa}.

The inifnitesimal generator matrix characterizing the ICTMC is denoted with $\mathbf{Q}(t)$, where the off-diagonal elements of $\mathbf{Q}(t)$ are the instantaneous transition rates between two different states in $\mathcal{S}$ at time $t$ and the diagonals are such that each row sum of $\mathbf{Q}(t)$ is $0$. 
The finite-time transition probability matrix of the ICTMC from time $t_0$ to $t$ is denoted with $\mathbf{P}(t_0, t)$.
If $\mathbf{Q}(t)$ is piecewise continuous, then the solution for $\mathbf{P}(t_0, t)$ is given by the Peano-Baker series 
\citep{Fortmann1977, Kailath1980}
\begin{equation}\label{eq:peano_baker2}
\begin{split}
\mathbf{P}(t_0, t) &= \mathbf{I} + \int_{t_0}^{t} \mathbf{Q}(\tau_1) d\tau_1 + \int_{t_0}^{t} \mathbf{Q}(\tau_1) d\tau_1 \int_{t_0}^{\tau_1} \mathbf{Q}(\tau_2) d\tau_2 d\tau_1 + \dots \\
& + \int_{t_0}^{t} \mathbf{Q}(\tau_1) \int_{t_0}^{\tau_1} \dots \int_{t_0}^{\tau_{k - 1}} \mathbf{Q}(\tau_k) d\tau_k d\tau_{k - 1} \dots d\tau_2 d\tau_1 + \dots ,
\end{split}
\end{equation}
which does not have a closed-form expression in general. 
However, it becomes tractable if $\mathbf{Q}(t)$ commutes with $\mathbf{Q}(t')$ for all $t$, $t'$.
This case naturally holds when $\mathbf{Q}(t) = f(t) \mathbf{Q}$, where $\mathbf{Q}$ is a valid infinitesimal generator matrix independent of time and $f(t)$ is an unknown, positive and integrable function of time.
Under this assumption, the transition probability matrix of the ICTMC acting along the branch connecting node $i$ to $\pa{i}$ can be expressed as a function of the integral of $f(t)$ from $t_{\pa{i}}$ to $t_{i}$ \citep{Rindos} as shown below
 \begin{equation}
 \mathbf{P}(t_{\pa{i}}, t_i) = \exp\left[\left(\int_{t_{\pa{i}}}^{t_i} f(t) dt \right) \mathbf{Q}\right]. 
 \label{eq:ictmc-tpm2}
\end{equation}
Here, $f(t)$ describes the trajectory of the evolutionary rate through time.
The likelihood of the observed data at site $c$ is the probability of the characters $\mathbf{Y}^{(c)} = (Y_{1,c}, \dots, Y_{N,c})'$ observed at the tips, marginalized over all possible latent states $\mathbf{Y}^{(c)*} = (Y_{N + 1,c}, \dots, Y_{2N - 1,c})$ at the internal nodes, given the model parameters $\boldsymbol{\theta}$ as follows
\begin{equation}
\begin{split}
\mathrm{P}(\mathbf{Y}^{(c)} \given \boldsymbol{ \theta}) &=  \sum\limits_{Y_{N+1,c}} \dots \sum\limits_{Y_{2N - 1,c}} \mathrm{P}(\mathbf{Y}^{(c)}, \mathbf{Y}^{(c)*} \given \boldsymbol{\theta}) \\
\mathrm{P}(\mathbf{Y}^{(c)}, \mathbf{Y}^{(c)*} \given \boldsymbol{\theta}) &= \mathrm{P}(Y_{2N - 1,c} | \boldsymbol{\theta}) \prod_{j = 1}^{2N - 2} \mathrm{P}(Y_{j,c} | Y_{\pa{j},c} , \boldsymbol{\theta})
\end{split}
\label{eq:likelihood}
\end{equation}
The likelihood can be computed efficiently using the pruning algorithm \citep{Felsenstein1973, Felsenstein1981}, which traverses the tree in a post-order fashion, proceeding from the tips towards the root, visiting each node exactly once, until the root node is reached. 
The likelihood for multiple sites can be expressed as a weighted mixture of the likelihood at each site under an across-site rate variation model.
The integral of $f(t)$ appears in the likelihood through the transition probabilities $\mathrm{P}(Y_{j,c} \given Y_{\pa{j},c}, \boldsymbol{\theta})$ for $j = 1, \dots, 2N - 2$.

\subsection{Spline Clock Model}
\label{subsec:scm}
To evaluate the integral of $f(t)$ over all $2N - 2$ branches of the phylogeny efficiently and to obtain a smooth and flexible estimate of the rate trajectory, we model $f(t) = h(g(t))$, where $h(\cdot)$ is a positive link function and $g(\cdot)$ is represented using B-splines   \citep{Schoenberg1946, deBoor1977, deBoor1978, Cox1982, Dierckx1993}.
B-splines are piecewise polynomial basis functions defined over a non-decreasing knot sequence $w_1 \leq \dots \leq w_{K}$ and characterized by their polynomial degree $d$ as follows:
\begin{equation*}
g(t) = \beta_0 + \sum_{i = 1}^{K + d} \beta_i B_{i + 1, d} (t),
\end{equation*}
where $B_{i, d}$ is the $i$-th B-spline basis function of degree $d$, constructed via the Cox-de Boor recursion \citep{deBoor1978}. 
$\beta_0$ is the intercept and $\beta_i$ is the coefficient associated with $B_{i, d}$ for $i = 1, \dots, K + d$.
We use $\boldsymbol{\beta} = (\beta_1, \dots, \beta_{K + d}) \transpose$ to collectively denote the coefficient parameters excluding the intercept.
Among splines of various degrees, 
a cubic B-spline ($d = 3$) defined on a strictly increasing sequence of knots is piecewise infinitely differentiable between the knots, and of continuity $C^2$ at each knot \citep{Goldman2002}, meaning it is continuous with continuous first and second derivatives at each knot, making it suitable for obtaining smooth estimates of unknown functions.
Therefore, for this work, we use cubic splines by fixing $d = 3$.
The number and placement of knots govern the flexibility of the approximation and their selection has received considerable attention in the literature. 
With sufficiently many knots, B-splines can approximate any continuous function on a compact interval to arbitrary accuracy \citep{Schumaker2007}.
Choices previously used in the literature include equally spaced knots \citep{Eubank1999, Schumaker2007} and quantiles of the observed covariate distribution \citep{Ruppert2003}.
\cite{Eilers1996} further popularized the use of a large number of knots combined with a difference penalty on adjacent B-spline coefficients to avoid overfitting, giving rise to the widely used P-splines framework. 
In this work, we use Bayesian P-splines \citep{LangBrezger2004, BrezgerLang2006} by 
placing a modest number of knots over the temporal domain and imposing a first order random walk prior on the B-spline coefficients for smoothing.

Next, we choose a suitable positive link function $h(\cdot)$ such that the integral of $h(g(t))$ over the branches can be evaluated efficiently.
A natural candidate is the exponential link $h(t) = \exp(t)$ which is positive and strictly increasing. 
However, the integral of $ \exp (g(t))$ appearing in the transition probability matrix of the ICTMCs does not admit a closed-form expression and needs to be approximated numerically.
The squared link $h(t) = t^2$ has the attractive property that the branch integrals admit closed-form expressions.
However, the squared link is non-monotonic, which can introduce identifiability problems, rendering the likelihood multimodal and the coefficients unidentifiable up to sign changes.
Although the quantity of primary interest, $f(t)=g(t)^2$, is itself identifiable even when the coefficients are not, the multimodality of the likelihood surface poses practical challenges such as poor mixing  across modes using MCMC algorithms in a Bayesian inference framework.
We further discuss why the prior imposed on $f(t)$ by the exponential link is more suitable than the prior imposed on $f(t)$ by the squared link (see Appendix~\ref{sec:appendixB} for details).
Therefore, we proceed with the exponential link which is the more standard choice used in the literature for log-linear modeling of hazard and intensity functions \citep{Fahrmeir2001, Wood2017, Rue2009}.
Under this parameterization, $g(t)$ is interpretable as the log-rate, which is a more natural and intuitive interpretation. 
We approximate the integral of $\exp(g(t))$ across all branches numerically using Gauss-Legendre (GL) quadrature \citep{Gauss1814}, a method that selects quadrature nodes and weights optimally in the sense that an $n$-point rule integrates any polynomial of degree at most $2n-1$ exactly, and achieves high accuracy for smooth integrands \citep{Press2007}.
Exploiting the piecewise polynomial structure of $g(t)$, we apply the GL rule locally on each knot interval that overlaps with the branch rather than over the entire branch at once.
On each such interval, $g(t)$ is a polynomial of degree $d$, so $\exp(g(t))$ is smooth and well-approximated by a polynomial of higher degree, making GL quadrature highly accurate.
Concretely, on a sub-interval $[L, R]$ of a knot interval, the 5-point GL rule approximates the integral as
\begin{equation}
     \int_L^R \exp(g(t))\,dt
     \approx \frac{R - L}{2}
     \sum_{j=1}^{5} a_j \exp\!\left(g\!\left(\frac{L+R}{2}
     + \frac{R-L}{2}\,x_j\right)\right)
     \label{eq:gl_quad},
\end{equation}
where $x_j$ and $a_j$ are the 5-point GL nodes and weights on $[-1,1]$ \citep{Press2007}.
The total branch integral is obtained by summing the contributions from all overlapping knot intervals.
This piecewise application of GL quadrature is exact for polynomials up to degree $9$ on each interval, and provides highly accurate approximations for the exponential link since the exponential of a cubic polynomial is smooth with well-controlled higher-order derivatives on any bounded interval.
A key advantage of this approach over adaptive quadrature methods such as the trapezoidal rule is that it incurs a deterministic, fixed computational cost which is exactly $5 \times (K + 2d + 1)$ function evaluations per overlapping knot interval, independent of the integrand shape.
Adaptive integrators, by contrast, refine their approximation iteratively until a convergence criterion is satisfied and can require substantially more function evaluations than the piecewise GL approach to achieve comparable accuracy, potentially leading to slow convergence or significantly higher runtime.
 
\subsection{Bayesian Inference}\label{sec:bayesian_inference2}

In a Bayesian framework, inference proceeds by sampling from the joint posterior density of all parameters given the observed data.
Besides the parameters $\beta_0$ and $\boldsymbol{\beta}$, the likelihood in \ref{eq:likelihood} depends on the tree topology and branch lengths $\mathcal{F}$, the across-site rate variation model parameter $\alpha$ and the substitution model $\mathbf{Q}$.
The joint posterior density of all the parameters given the data is proportional to the product of the data likelihood and the joint prior density 
\begin{equation}
\mathrm{P}(\beta_0, \boldsymbol{\beta}, \mathbf{Q}, \alpha, \mathcal{F} \given \mathbf{Y}) \propto \mathrm{P}(\mathbf{Y} \given \beta_0, \boldsymbol{\beta}, \mathbf{Q}, \alpha, \mathcal{F}) \mathrm{P}(\beta_0) \mathrm{P}(\boldsymbol{\beta})  \mathrm{P}(\mathbf{Q}) \mathrm{P}(\alpha) \mathrm{P}(\mathcal{F}).
\label{eq:jointposterior2}
\end{equation}
A key modeling choice is the prior placed on the primary parameters of interest, $\beta_0$ and  $\boldsymbol{\beta}$. 
A commonly used approach is Bayesian P-splines \citep{LangBrezger2004}, under which a Gaussian Markov random field (GMRF) \citep{rueheld} prior is assigned to $\boldsymbol{\beta}$ as follows
\begin{equation}
\beta_{i + 1} - \beta_i \overset{\text{ind}}{\sim} \mathcal{N}(0, 1/\tau),
\end{equation}
for $i = 1, \dots, K + d - 1$.
This avoids overfitting by penalizing the roughness in the estimated curve and shrinking the differences between adjacent coefficients toward zero, thereby controlling the smoothness of $f(t)$ through the prior rather than by limiting the number of knots.
This is equivalent to assigning the following joint prior to $ \boldsymbol{\beta}$
\begin{equation}
\mathrm{P}( \boldsymbol{\beta} \given \tau) \propto \tau^{\frac{K + d - 1}{2}} \exp{-\left[\frac{\tau}{2} \boldsymbol{\beta}' \left(\mathbf{D} - \mathbf{A}\right) \boldsymbol{\beta}\right]},
\label{eq:weighted_gmrf3}
\end{equation}
where $\mathbf{A} = \left\{a_{ij}\right\}$ is the adjacency matrix with entries defined as $a_{ij} = 1$ if $|i - j| = 1$ and $0$ otherwise and $\mathbf{D}$ is a diagonal matrix where the $i$-th diagonal entry of $\mathbf{D}$ is equal to the $i$-th row sum of $\mathbf{A}$. 
The precision matrix $\mathbf{D} - \mathbf{A}$ is rank deficient and therefore the prior is improper. 
This necessitates verifying that the resulting posterior is proper in order to sample from the posterior using MCMC algorithms.
In many settings, improper priors yield proper posteriors because the likelihood provides sufficient regularization by decaying in the tails of the parameter space.
However, in our case, the likelihood of the observed sequence data does not converge to $0$ as $\boldsymbol{\beta} \rightarrow \infty$ but instead converges to a positive constant.
The impropriety can be remedied by redefining the precision matrix of the prior  as $\mathbf{D} - \rho \mathbf{A}$, where $|\rho| < 1$, giving rise to a class of proper GMRF priors.
We proceed by assigning the proper GMRF prior with $\rho$ fixed at $0.99$ to $\boldsymbol{\beta}$ in order to incorporate smoothing of the estimated rate function while maintaining posterior propriety.
We assign a $\mathcal{N}(0, 2)$ prior to $\beta_0$ and a Gamma prior with shape $4$ and scale $2$ to the GMRF precision $\tau$.
Additionally, we assign suitable priors $\mathrm{P}(\mathbf{Q})$, $\mathrm{P}(\alpha)$ and $\mathrm{P}(\mathcal{F})$ to $\mathbf{Q}$, $\alpha$ and $\mathcal{F}$ respectively following standard Bayesian phylogenetic practices \citep{Baele2025}. 
We sample from the joint posterior density in \ref{eq:jointposterior2} within a Metropolis within Gibbs scheme.
See  \citet{Drummond2002}, \citet{Drummond2012} and \citet{Fisher2022} for details on proposal distributions used over phylogenetic random variables like $\mathcal{F}$, $\mathbf{Q}$ and $\alpha$.
In this work, we pay special attention to the conditional sampling of the spline coefficients $\beta_0$ and $\boldsymbol{\beta}$, which we accomplish via Metropolis–Hastings \citep{Metropolis1953, Hastings1970} steps with random-walk proposals.

\section{Results}
\label{sec:Results}
 
\subsection{Simulation Study}
\label{subsec:Simulation}

\begin{figure}[htbp]
     \centering
   	\includegraphics[width=0.95\linewidth]{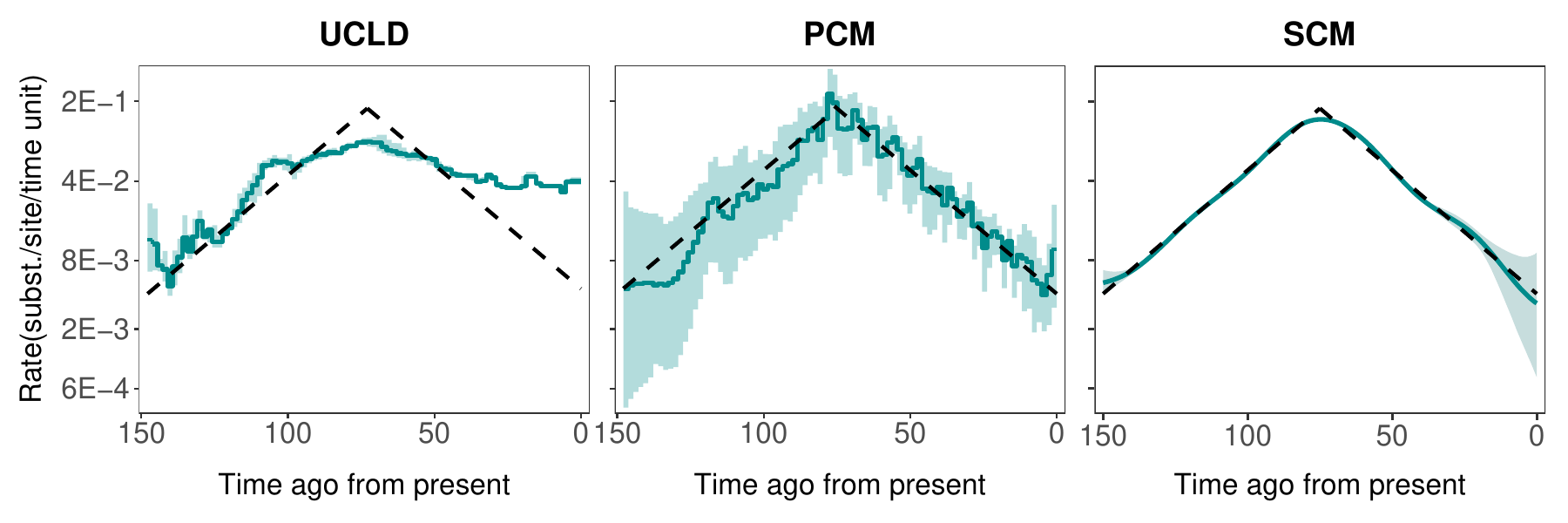} 
  	\caption{Simulation results.
  	The posterior median (solid green line) and the $95\%$ Bayesian credible interval or BCI (green shaded area) of the evolutionary rate are shown for the uncorrelated relaxed clock model with lognormal distribution (UCLD), the polyepoch clock model (PCM) and the spline clock model (SCM). The true rate is  indicated with a black dashed line.}
  	\label{fig:simulation}
\end{figure}

We examine the ability of the spline clock model to recover the true evolutionary rate from sequence data simulated under time-varying rates.
We use a tree of height 160 time units with $100$ tip nodes simulated from the exponential growth coalescent model with initial population size $10,000$ and growth rate $0.05$. 
We use $\pi$BUSS \citep{pibuss} to simulate sequences of length $10,000$ along the tips of the simulated tree under an HKY substitution process \citep{hky1985} with a fixed transition/transversion ratio of 2 and a time-dependent rate that initially increases log-linearly before declining sharply
 \[
 f(x) =
 \begin{cases}
 \exp(-5.5 + 0.05 x) & \text{if } x \geq 75 \\
 \exp(2 - 0.05 x) & \text{if } x < 75.
 \end{cases}
 \]
We first analyze the simulated data under the uncorrelated relaxed clock with underlying log-normal distribution \citep{Drummond2006}.
The uncorrelated relaxed clock is one of the most commonly used relaxed clock models that allows each branch of the phylogenetic tree to have its own evolutionary rate,  which does not depend on the rate associated with the neighboring branches, and the commonly used assumption is that the the rates follow a log-normal distribution.
Under this model, the heterogeneity in rate is with respect to the branches and not with respect to time.
Therefore, we estimate the posterior median, $2.5\%$ and $97.5\%$ percentiles of the evolutionary rate for extant branches across a regular grid of fixed times.
For each tree, we average the branch specific rates of the branches that pass through each time point to get an average rate at that point for each tree.
At each point, we consider the median of the average rates obtained at that point across all trees as the median of the rate at that point and we use the $2.5\%$ and $97.5\%$ quantiles of the average rate at each point across all trees to quantify the uncertainty of the rate at that point.
We also analyze the simulated data using the polyepoch clock model  \citep{Datta2025} with 100 epochs between 0 and 170.
Finally, we fit the spline clock model to the simulated data, placing knots at intervals of 20 time units between 0 and 160.
We fit each of the three clock models to the simulated data along with a non-parametric coalescent tree prior \citep{Gill2012} and the HKY substitution model \citep{hky1985}. 
We use our implementation of the spline clock model and implementations of the uncorrelated relaxed clock model and the polyepoch clock model in BEAST X \citep{Baele2025} to jointly infer the rate, tree topology, branch lengths, and other model parameters from the simulated sequence data. 
Figure~\ref{fig:simulation} presents the posterior inference results from fitting the three clock models to the simulated data set. 
The posterior median under the spline clock model closely tracks the true rate trajectory throughout the entire time range. 
While the post-hoc rate estimate under the uncorrelated relaxed clock initially reflects the increasing trend, it ultimately gets modulated and fails to recover the true rate, suggesting that the model lacks the flexibility to accommodate systematic temporal rate variation. 
The posterior median under the polyepoch clock model broadly follows the true rate, however, the estimate exhibits notable roughness attributable to discontinuities introduced by the discretization of the temporal domain, and the associated credible intervals are also comparatively wider. 
The spline clock model produces a smooth and accurate estimate with tighter credible intervals, demonstrating its suitability for capturing continuously varying evolutionary rates.
We also perform additional simulations under a constant rate and a strictly increasing rate (see Appendix~\ref{sec:appendixA} for details.)

\subsection{Exemplars}
\label{subsec:exemplars}

We apply the spline clock model to two real-world pathogen examples.
The first is a foamy virus (FV) alignment \citep{aiewsakun2015b} of 14 \textit{pol} sequences, used to examine deep evolutionary rate variation over approximately 100 million years. 
The second exploits the flexibility of CTMC models beyond nucleotide substitution. 
The same mathematical framework applies to any discrete trait evolving along a phylogeny, including geographic location. 
Under this formulation, pathogen dispersal across discrete locations is modeled as a CTMC along the branches of a timed phylogeny, enabling simultaneous inference of evolutionary history and spatial diffusion \citep{Lemey2009}. 
In epidemic settings, transmission dynamics change over time due to public health interventions  and shifts in epidemic waves. 
Replacing the homogeneous CTMCs with ICTMCs whose diffusion rate varies smoothly over time allows more realistic characterization of the spatio-temporal spread. 
We exploit this capacity by fitting the spline clock model to 3,959 SARS-CoV-2 genomes \citep{Lemey2021} from 10 European countries collected in 2020, inferring temporal variation in the rate of spatial diffusion during the first and second epidemic waves.
For all analyses, we run MCMC chains for 100 million iterations, discarding 10\% of samples as burn-in and thinning every 2,000 iterations.
We assess convergence by verifying that the effective sample size (ESS) is greater than 500 for all parameters of interest in Tracer \citep{Tracer}. 
We summarize the posterior distribution over tree topologies using the maximum clade credibility (MCC) tree, which has the maximum product of the posterior clade probabilities, providing a single representative topology annotated with marginal posterior summaries of node heights. 
The MCC tree is  inferred in TreeAnnotator~X \citep{HIPSTR}.
We provide instructions and the BEAST XML files for reproducing these analyses on Github at \url{https://anonymous.4open.science/r/spline_clock_model_supplement-1376}.

\subsubsection{Foamy Virus}
\label{subsubsec:foamyVirus}

\begin{figure}[htbp]
     \centering
   	\includegraphics[width=0.65\linewidth]{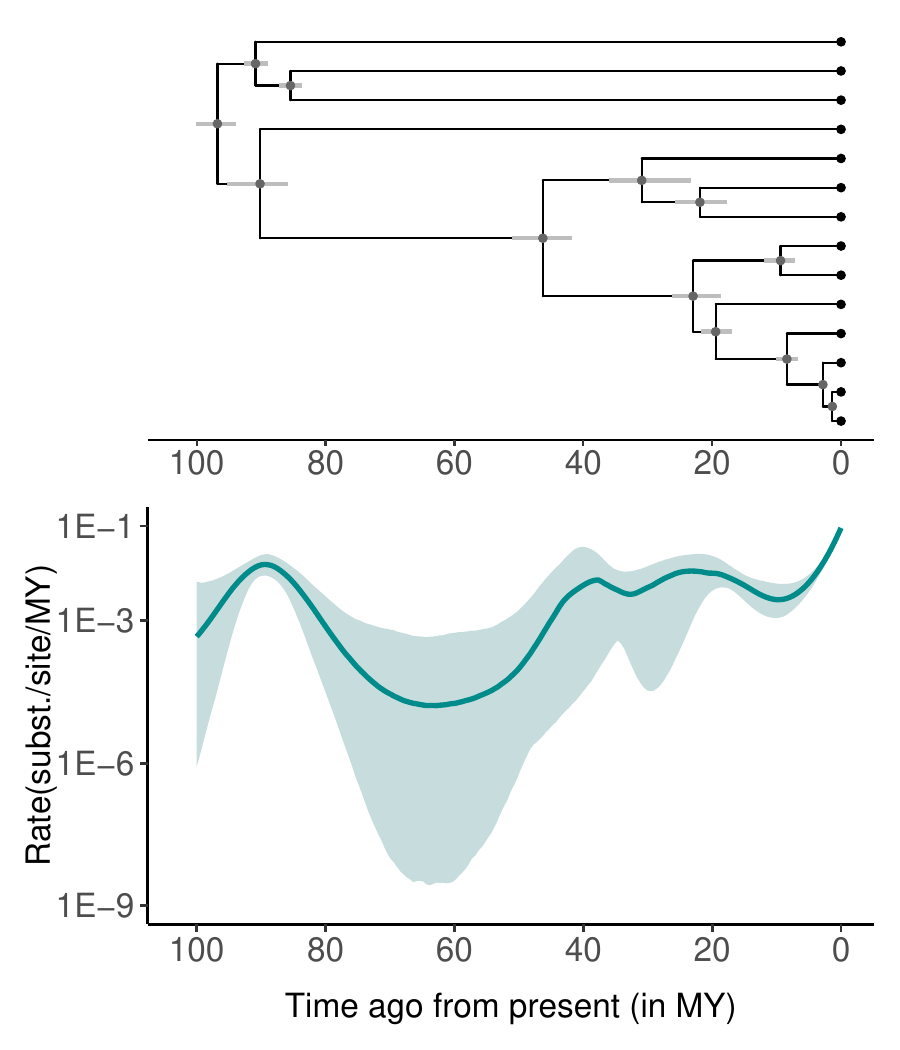} 
  	\caption{Evolutionary dynamics of the foamy virus over past 100 million years (MY).
  	The top panel shows the inferred maximum clade credibility (MCC) tree, where the posterior median and $95\%$ BCI of the internal node dates are depicted with gray circles and gray shaded bars respectively.
  	The bottom panel shows the posterior median (solid green line) and the $95\%$ BCI (green shaded area) of the evolutionary rate.
  	}
  	\label{fig:fv}
\end{figure}

FVs are a group of complex retroviruses that have a very stable and long co-speciation history with their hosts, stretching back more than a hundred million years \citep{Switzer2005, Katzourakis2009, Katzourakis2014}.
As a result, the divergence times of most FV lineages can be inferred directly from the divergence history of their hosts \citep{BinindaEmonds2007, Stone2010, Perelman2011}.
The data set consists of an alignment of 14 FV \textit{pol} sequences (3,351 nt) from several primates, one bovine, one equine, and one feline host. 
\cite{aiewsakun2015b} and \cite{Membrebe2019} previously analyzed this data set and demonstrated a discrepancy between long-term and short-term rate estimates, which they attributed to a power-law relationship between rate and time.
Here, we analyze the FV data set under the spline clock model to obtain a smooth estimate of the rate trajectory without imposing a strong parametric relationship between rate and time.
We fit the spline clock model with the Yule speciation tree prior \citep{Yule1925}, the general time-reversible substitution model \citep{Tavare1986} and rate heterogeneity among sites modeled using a discretized gamma distribution \citep{Yang1996}.
We use host divergence time estimates as calibrations for the FV cospeciation history, and place knots at intervals of 10 million years (MY) between 100 million years ago (MYA) and 0 MYA, based on time of most recent common ancestor (TMRCA) estimates from prior studies \citep{aiewsakun2015b, Membrebe2019}, to ensure the spline has sufficient resolution to capture rate variation across the full temporal range of the data.
Figure~\ref{fig:fv} shows the posterior inference results obtained for the FV data analysis.
The top panel shows the MCC tree, with median TMRCA 98 MYA (96-101).
The bottom panel shows the posterior median of the rate (solid green line) and the $95\%$ BCI (green shaded area).
We observe a strong time varying effect in the rate estimate, with variations over at least four orders of magnitude.
Closer to the root, the estimated rate is $ 3.71 \times 10^{-4}$ ($1.75 \times 10^{-6}$ - $6.32 \times 10^{-3}$) substitutions (subs) / site / MY, which then increases to $1.6 \times 10^{-2}$ ($9.32 \times 10^{-3}$ - $2.58 \times 10^{-2}$) subs / site / MY around $90$ MYA. 
Here ranges within the parenthesis denote 95\% Bayesian credible intervals (BCI).
From 90 MYA to 60 MYA, the rate drops to $1.41 \times 10^{-5}$ $(4.46 \times 10^{-9} - 4.39 \times 10^{-4})$ subs / site / MY and increases back again by 40 MYA.
We observe an increased uncertainty in the rate estimate during this time period due to the lack of calibration information.
Between 10 MYA to 0 MYA, we get a more precise rate estimate exhibiting an increase from $3.29 \times 10^{-3}$ ($1.30 \times 10^{-3}$ - $7.31 \times 10^{-3}$) subs / site / MY to $9.22 \times 10^{-2}$ ($95\%$ BCI $7.16 \times 10^{-2}$ - $1.24 \times 10^{-1}$) subs / site / MY.
The estimated rate near the present is substantially higher than the estimate 60 MYA, highlighting the importance of accounting for rate variation over deep evolutionary timescales.  
However, the estimated rate trajectory does not follow the power-law decay pattern typically associated with the time-dependent rate phenomenon \citep{aiewsakun2015b, Membrebe2019}. 
We observe non-monotonic variation with elevated rates near the root.
\citet{Membrebe2022} extended the power-law rate decay epoch model with branch-specific random effects and applied it to the FV data, recovering the largest effect for a short ancestral branch near the root, which is consistent with the elevated rate estimate near the root we recover here. 
This pattern may partly reflect among-lineage rate heterogeneity in addition to rate variation through time and naturally motivates extending the spline clock model to jointly capture rate variation across both time and lineages, with the ICTMC framework offering a principled foundation for such an extension.

\subsubsection{SARS-CoV-2}
\label{subsubsec:covid}

The COVID-19 pandemic caused by SARS-CoV-2 claimed an estimated 7 million lives globally. 
In Europe, the first wave of infections was successfully contained in Spring 2020 gradually leading to relaxation of containment measures.
However, cases began surging again across much of Europe in late Summer 2020, culminating in a second wave by October 2020.
We analyze 3959 SARS-CoV2 genomes from Belgium, France, Germany, Italy, the Netherlands, Norway, Portugal, Spain, Switzerland and the UK, from both the first and second wave, collected from GISAID on November 3, 2020 \citep{Lemey2021}.
We examine the temporal variation in the rate of spatial diffusion of  SARS-CoV-2 between these European countries.
We perform a joint Bayesian inference of sequences and traits by integrating genome sequences with  sampling location and date, as well as mobility and connectivity data.
We model the process of transitioning through discrete location states, which are the countries of sampling in this case, using the spline clock model, and we model the relative rates as log-linear functions of mobility and connectivity covariates following the original study \citep{Lemey2021}. 
We model the effective population size trajectory assuming a piecewise-constant function \citep{Gill2012}, where the log-transformed population sizes are modeled as a deterministic function of log transformed counts of cases of COVID-19 over 2 week intervals and the covariates include social connectedness index (SCI) of Facebook, air transportation data and mobility data.
The inclusion probability of each potential covariate is estimated through a spike-and-slab procedure \citep{Lemey2014}.

\begin{figure}[htbp]
     \centering
   	\includegraphics[width=0.7\linewidth]{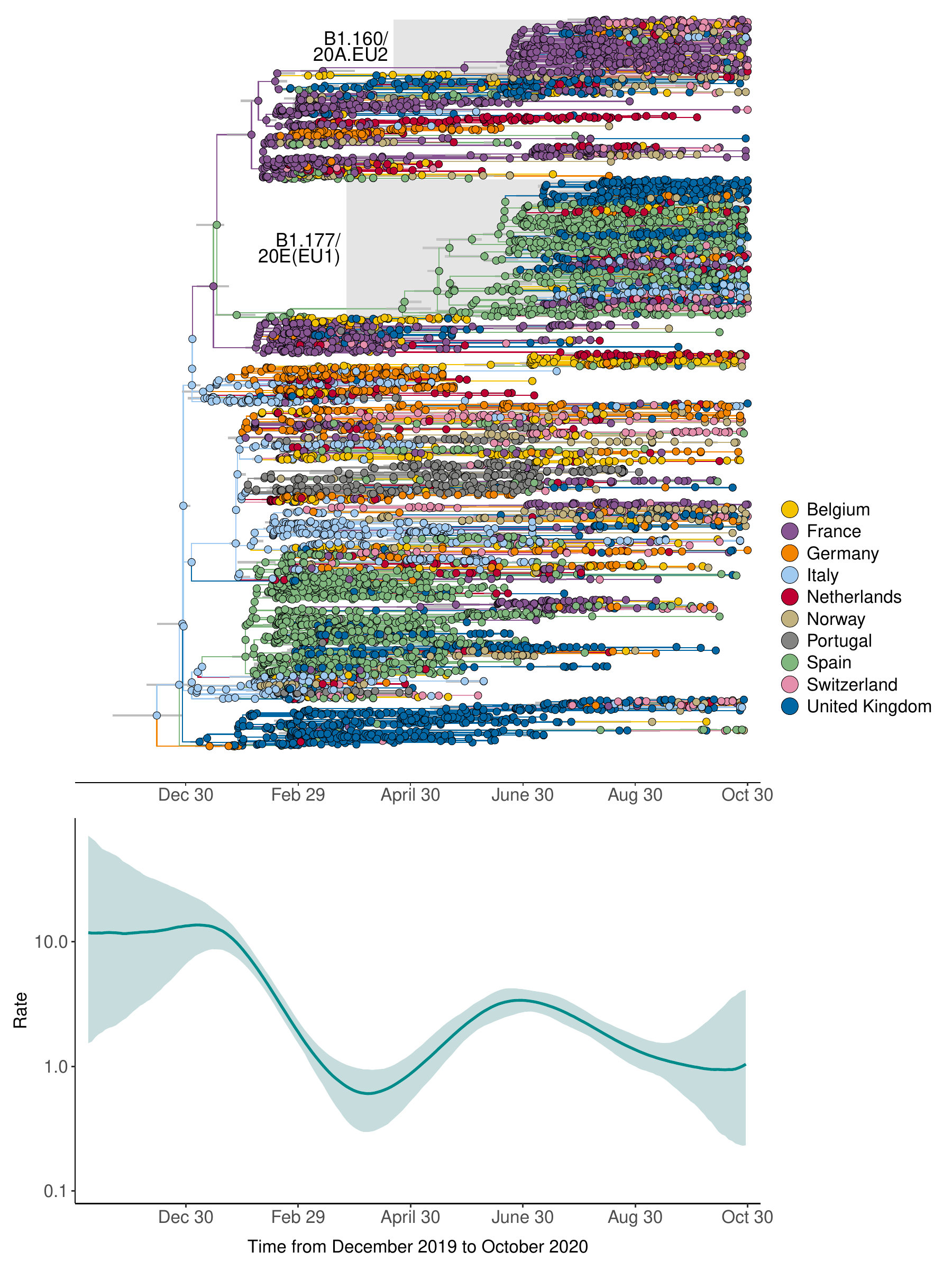} 
  	\caption{SARS-CoV-2 in Europe from December 2019 to October 2020.
  	The top panel shows the inferred MCC tree.
  	Nodes and branches are colored according to the countries.
  	The bottom panel shows the posterior median (solid green line) and the $95\%$ BCI (green shaded area) of the rate of spatial diffusion.
  	}
  	\label{fig:covid}
\end{figure}

Figure~\ref{fig:covid} shows the posterior inference results.
The bottom panel shows the posterior median and $95\%$ BCI of the rate of spread of SARS-CoV-2 between the 10 European countries.
We observe that the rate decreases from 10 to less than 1 between January and April, coinciding with the imposition of lockdown measures to curb transmission during the first wave.
Having successfully contained the first wave in spring 2020, containment measures were gradually being relaxed by mid-April 2020 and starting from around this time, we observe an increase in the rate of spread, reaching a peak in July.
This aligns with the onset of the second wave as infections starting rising rapidly in late summer leading to reimposition of lockdown and we observe a decrease in the rate of spread from July to October.
The top panel of Figure~\ref{fig:covid} shows the inferred MCC tree, where nodes and branches are colored according to the countries.
We observe that the TMRCA is December 2019.
We also identify the variants B.1.160/20A.EU2 and B.1.177/20E(EU1), indicated with highlighted clades.
Our findings support Spain as the origin of variant B.1.177/20E(EU1), which subsequently spreads to become the predominant strain in the UK by late summer 2020, in agreement with \cite{Lemey2021}.
Additionally, variant B.1.160/20A.EU is identified as having disseminated primarily from France over the course of summer 2020.

\section{Discussion}
 \label{sec:Discussion}
 
Motivated by accumulating evidence for the time-dependent nature of evolutionary rates \citep{Aiewsakun2016, Duchene2014, aiewsakun2015b, Membrebe2019, Datta2025}, we develop the spline clock model, a novel and flexible approach to obtain smooth estimates of the rate trajectory through time from molecular sequence data.
The spline clock model accommodates time-varying rates by modeling the underlying process as an ICTMC acting along the branches of a timed phylogeny, with the log transformed rate parameterized flexibly using a cubic B-spline basis expansion.
This yields rate estimates that vary continuously and smoothly over the entire temporal domain.
The transition probability matrix of the ICTMC along each branch reduces to a function of the integral of the rate function over that branch \citep{Fortmann1977, Kailath1980, Rindos}, which needs to be computed for all $2N - 2$ branches of the phylogeny. 
Although, these integrals do not admit closed-form expressions, owing to the smoothness of the rate function under the spline clock model, we can numerically approximate the integrals using  Gauss-Legendre quadrature \citep{Gauss1814, Press2007} applied locally on each knot interval. 
This approach exploits the piecewise polynomial structure of the log-rate function and achieves high accuracy while using fixed, deterministic function evaluations per branch.
To control the smoothness of the estimated rate, we use the Bayesian P-splines framework \citep{LangBrezger2004, BrezgerLang2006} and assign a proper GMRF prior \citep{rueheld} to the spline coefficients, where the GMRF precision is assigned a hyperprior and estimated from the data, providing a fully data-driven regularization scheme.

We assess the performance of the spline clock model by applying it to sequence data simulated under a piecewise log linear rate with an initial increase followed by a crash. 
The spline clock model successfully recovers the true rate trajectory, yielding a smooth posterior median that tracks the true rate more closely and with tighter credible intervals throughout the temporal domain, compared to competing clock models.
We further apply the spline clock model to two exemplars.
For the foamy virus, we recover strong temporal variation in the evolutionary rate spanning over four orders of magnitude across approximately 100 million years \citep{aiewsakun2015b, Membrebe2019}. 
We infer a notably elevated rate estimate near the root, which may be attributed to variation in rate across lineages in addition to variation through time. 
This motivates the development of more complex models that can simultaneously capture both sources of variation as a potential direction of future work.
Beyond evolutionary rates, the spline clock model extends naturally to other time-varying processes, such as the rate of spatial diffusion in phylogeography analyses. 
For SARS-CoV-2, the inferred temporal variation in the spatial diffusion rate aligns closely with real-world public health interventions, with marked decreases during lockdowns and a pronounced increase that  coincides with the relaxation of containment measures.
The spline clock model offers a principled way to account for systematic temporal variation in the evolutionary rate and fills a methodological gap between rigid parametric clock models and computationally prohibitive fully nonparametric alternatives, and we anticipate that it will prove broadly useful for uncovering temporal rate heterogeneity across a wide range of evolutionary settings.

\section*{Acknowledgement}
This work was supported through National Institutes of Health grants R01 AI153044 and R01 AI162611.
We gratefully acknowledge support from Advanced Micro Devices, Inc.~with the donation of parallel computing resources used for this research.
PL acknowledges support by the Research Foundation - Flanders (‘Fonds voor Wetenschappelijk Onderzoek - Vlaanderen’, G010326N and G051322N)

\bibliography{basis_approximation.bib}

\begin{appendices}

\setcounter{figure}{0}
\renewcommand{\thefigure}{A\arabic{figure}}

\section{Additional Simulations}
\label{sec:appendixA}

\begin{figure}
    \centering
  	\includegraphics[width=1.0\linewidth]{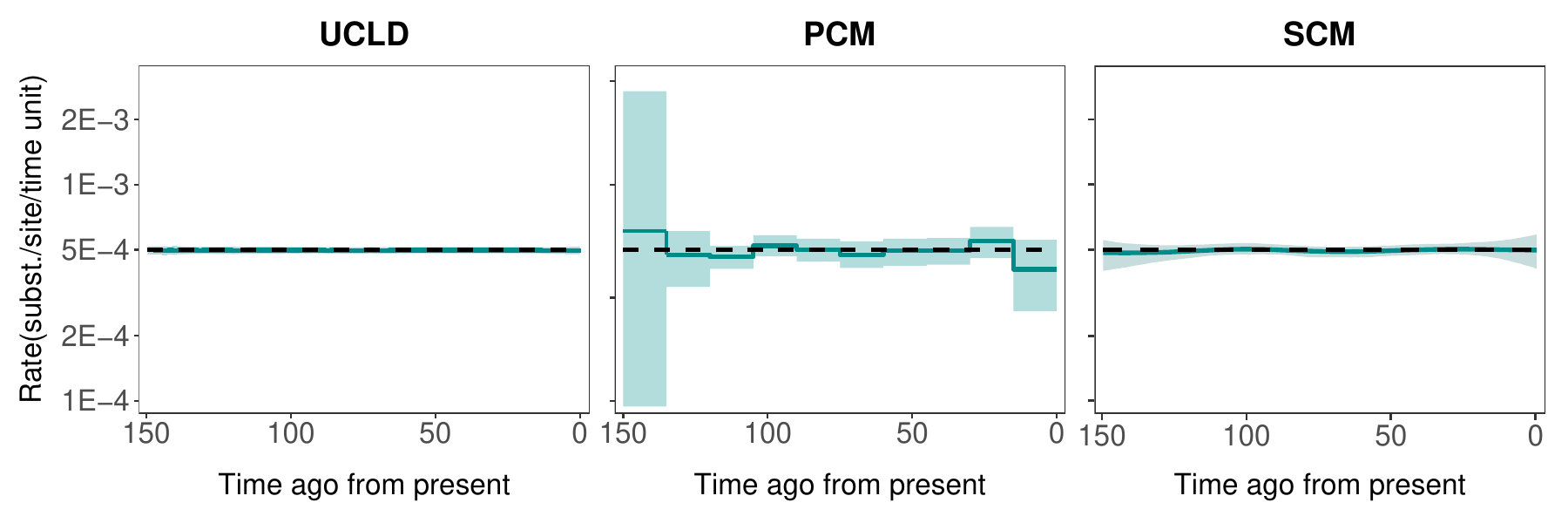} 
 	\caption{Constant rate simulation results.
 	The posterior median (solid green line) and the $95\%$ BCI (green shaded area) of the evolutionary rate are shown for the uncorrelated relaxed clock model (UCLD) on the left, the polyepoch clock model (PCM) in the center, and the spline clock model (SCM) on the right.}
 	\label{fig:scrsim2}
\end{figure}

\begin{figure}
    \centering
  	\includegraphics[width=1.0\linewidth]{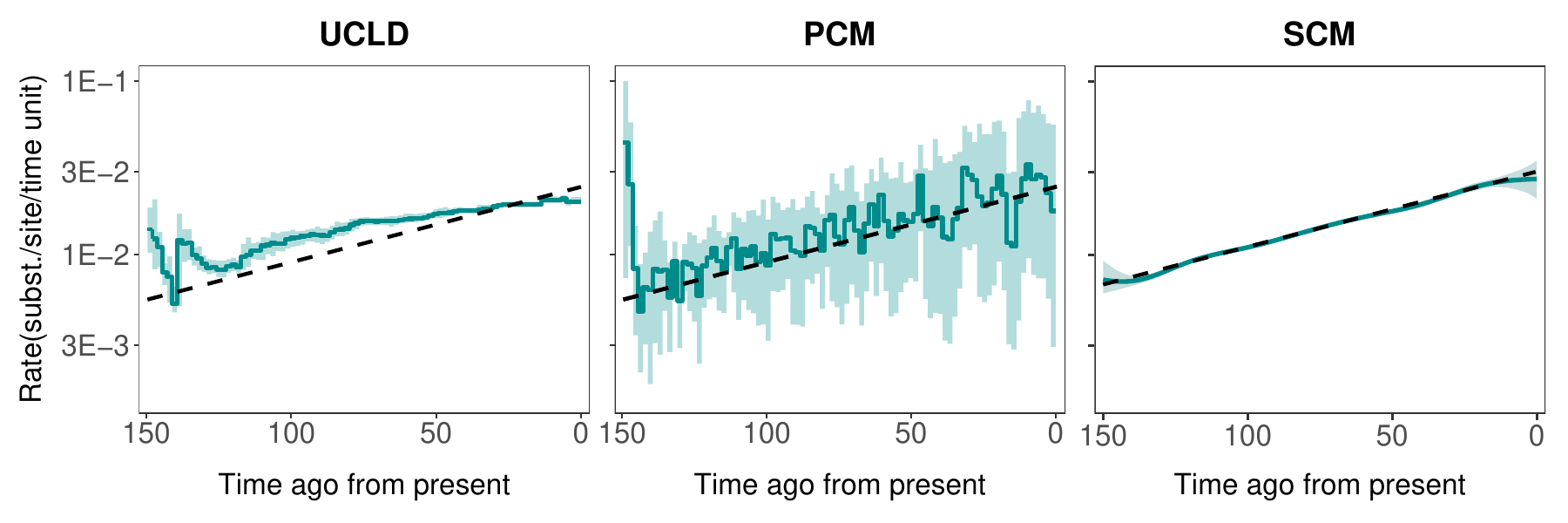} 
 	\caption{Increasing log-linear simulation results.
 	The posterior median (solid green line) and the $95\%$ BCI (green shaded area) of the evolutionary rate are shown for the UCLD on the left, the PCM in the center, and the SCM on the right.
 	The black dashed line shows the true rate.}
 	\label{fig:loglinsim2}
\end{figure}

We perform additional simulations to assess the performance of the spline clock model relative to the uncorrelated relaxed clock model with an underlying lognormal distribution and the polyepoch clock model, under two additional simulation scenarios.
First, we simulate sequences of length 10,000 under a constant rate of $5 \times 10^{-4}$ substitutions (subs) / site / year along the 100 tips of a tree of height 160 time units simulated under an exponential growth coalescent model with initial size 10,000 and growth rate 0.05.
We fit the polyepoch clock model with 10 epochs between 0 and 150.
We fit the spline clock model with the knots placed 20 time units apart between 0 and 160.
Figure~\ref{fig:scrsim2} shows the posterior inference results obtained by analyzing the sequence data simulated under a constant rate under the three different clock models.
We observe that the post-hoc rate estimate obtained using the uncorrelated relaxed clock model (see Section 3.1 of the main manuscript for details) as well as the posterior median of the rate obtained using the spline clock model align very closely with true rate throughout the entire temporal domain.
The posterior median under the polyepoch clock model is close to the true rate in general, but slightly deviates from the true rate, particularly closer to the present and the root.
Also, we observe higher uncertainty in the rate estimate under the polyepoch clock model compared to the other two models closer to the root.
Next, we simulate sequences under an increasing log-linear evolutionary rate $f(t) = \exp (-3.5 + 0.01 t)$. 
We fit the polyepoch clock model with 100 epochs between 0 and 160 to get a more flexible estimate. 
We use the same knot structure for the spline clock model as in the previous simulation.
Figure~\ref{fig:loglinsim2} shows the posterior inference results obtained by analyzing the sequence data simulated under the increasing log-linear rate under the three different clock models.
We observe that the posterior median of the rate obtained under the spline clock model nearly perfectly matches the true rate.
The post-hoc rate estimate obtained under the uncorrelated relaxed clock model gets modulated and does not align quite well with the true rate.
The posterior median of the rate estimate obtained using the polyepoch clock model has a general trend that aligns with true rate, however it is mixed with oscillations that deviate from the true log-linear rate.
Also, the the uncertainty in the estimates under the polyepoch clock model is considerably higher compared to the other two models.
Therefore, we observe that the spline clock model yields a smooth estimated rate trajectory that accurately captures the true evolutionary rate and yields tighter credible intervals compared to the competing clock models across both simulation scenarios.

\section{Squared Link}
\label{sec:appendixB}
\setcounter{figure}{0}
\renewcommand{\thefigure}{B\arabic{figure}}
We fit the spline clock model with the squared link to the data set simulated under the piecewise log-linear rate with initial increase followed by a crash (see Section 3.1 of the main manuscript for details on the simulation setup and knot structure).
Figure~\ref{fig:sqlink} shows the posterior inference results.
The top panel displays the posterior median and $95\%$ BCI of the estimated rate trajectory alongside the true rate. 
The posterior median deviates from the true rate closer to the present and we observe that the $95\%$ BCIs also become wider closer to the present, with the lower bound of the credible intervals dropping considerably.
\begin{figure}[t]
    \centering
   	\includegraphics[width=0.7\linewidth]{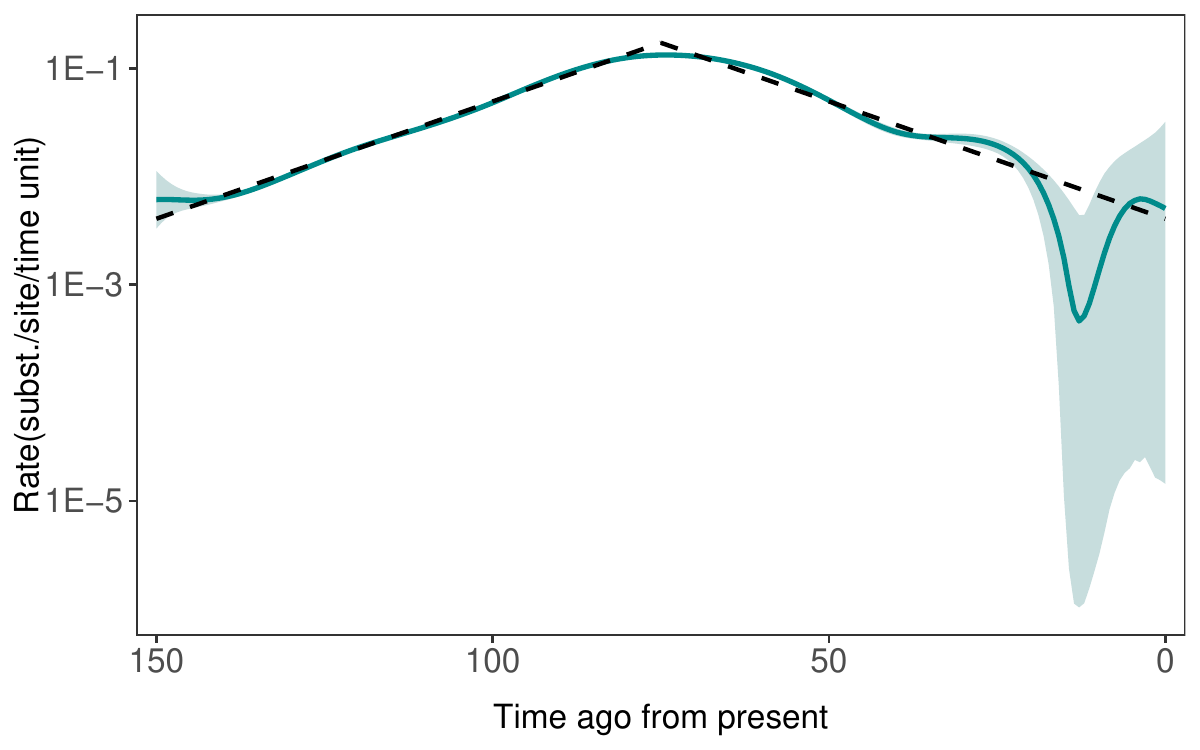} 
  	\caption{Piecewise log-linear simulation results using the spline clock model with squared link.
  	The posterior median (solid green line) and the $95\%$ BCI (green shaded area) of the evolutionary rate are shown along with true rate indicated with the black dashed line.}
  	\label{fig:sqlink}
\end{figure}
This could be attributed to poor mixing as well as to the prior induced on $f(t)$ when using the squared link.
First, we compare the mixing of the values of $f(t)$ at certain time points using the squared link with the mixing of $f(t)$ at the same time points using the exponential link.
Figure~\ref{fig:trace} shows the trace plot of $f(t)$ at the time points 10, 15 and 20 time units ago from present, which fall within the interval where the posterior median deviates from the true rate and the credible intervals get wider in Figure~\ref{fig:sqlink}.
We observe that the mixing is considerably better using the exponential link.
The effective sample size (ESS) per hour is 4, 2 and 1 at $t = 10, 15$ and 20 using the squared link.
The ESS per hour using the exponential link at these time points are 39, 109 and 74, which are considerably higher than the ESS per hour using the squared link.
These results show that the mixing of $f(t)$ is poor when we use the squared link.
\begin{figure}
     \centering
    	\includegraphics[width=0.9\linewidth]{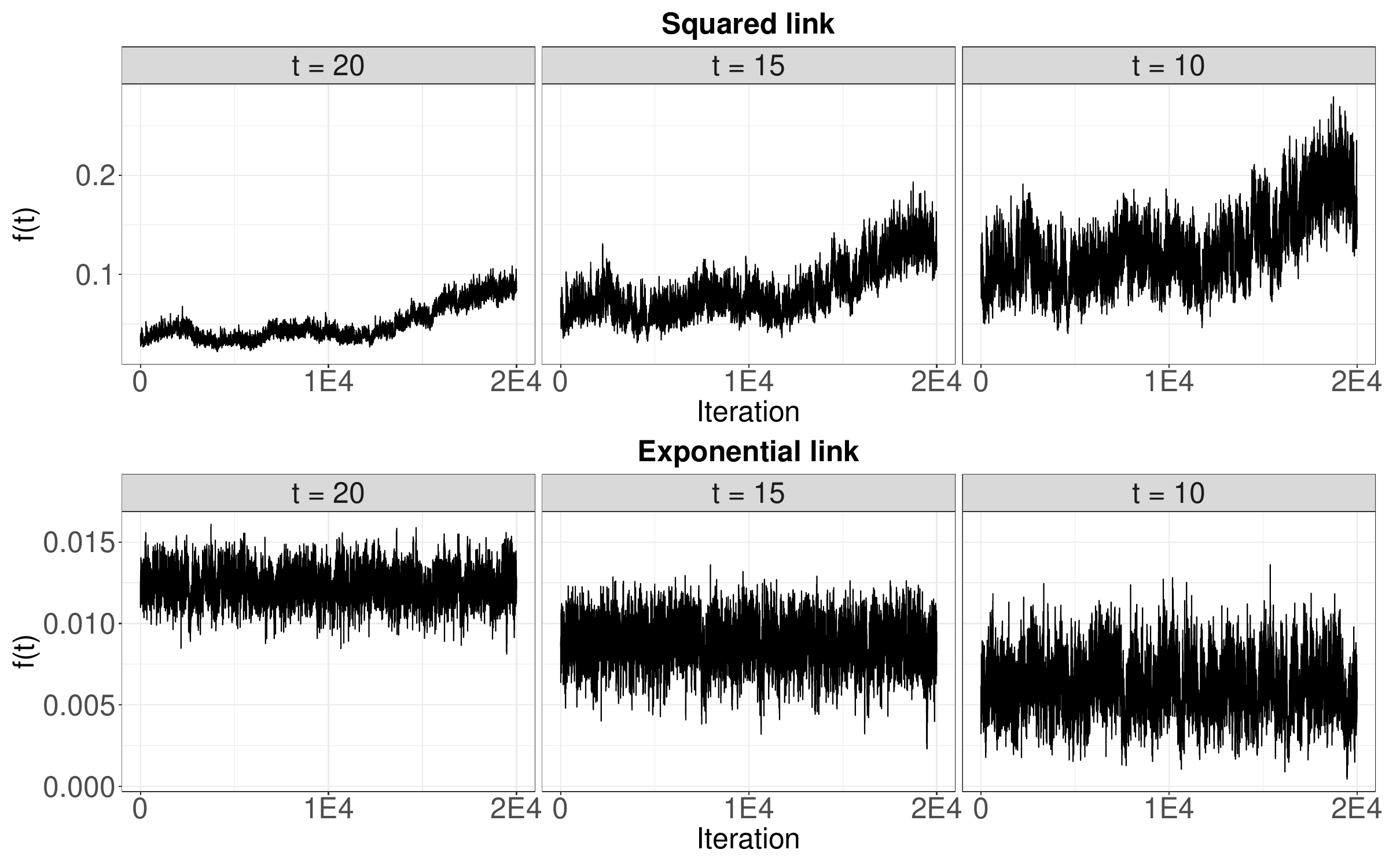} 
   	\caption{Trace plot of $f(t)$ at $t = 10, 15$ and 20 time units ago from present, using squared link and exponential link.}
   	\label{fig:trace}
\end{figure}
Next, we assess the difference in the prior induced on $f(t)$ for the two different link functions.
We draw 1000 samples of $(\beta_0$, $\boldsymbol{\beta}$ and $\tau)$ from the priors assigned to these parameters (see Section 2.3 of the main manuscript).
The knot structure is the same as the one used for fitting the spline clock model to the simulated data (see Section 3.1 of the main manuscript for details).
Figure~\ref{fig:prior_rate} shows the log transformed values of $f(t)$ using the squared and exponential link between $t = 150$ and 0 time units ago from present for each replicate of $(\beta_0, \boldsymbol{\beta}, \tau)$.
We observe that under the exponential link, the curves appears to be concentrated and symmetrically distributed around the horizontal line with intercept 0.
In contrast, under the squared link, the curves are more dispersed and exhibit a strong asymmetric distribution, with a pronounced tendency towards large negative values of log $f(t)$. 
Consequently, the prior under the squared link places substantial mass on rate trajectories that sharply drop close to zero and increase back sharply as well.
This is the exact pattern we observe in the posterior median near the present in Figure~\ref{fig:sqlink} and this also explains why the lower bound of the credible intervals drop sharply as well near the present.
This confirms that given the priors assigned to the spline clock model parameters, the resulting prior induced on $f(t)$ upon using the squared link is not suitable for estimating the unknown rate trajectory.
\begin{figure}[H]
     \centering
    	\includegraphics[width=0.9\linewidth]{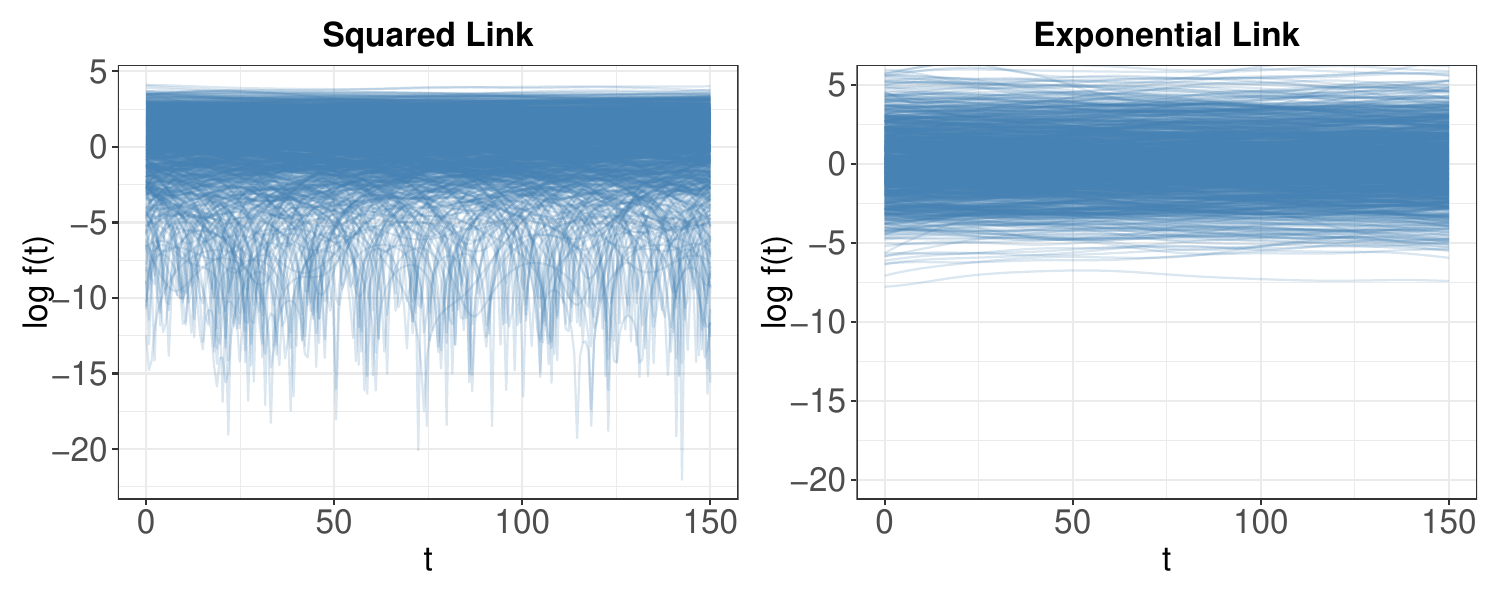} 
   	\caption{$\log f(t)$ curves using the squared and exponential link functions for 1000 samples of $\beta_0$, $\boldsymbol{\beta}$ and $\tau$ drawn from the assigned prior distributions.}
   	\label{fig:prior_rate}
\end{figure}
\end{appendices}

\end{document}